\documentclass[12pt]{article}
\usepackage{epsf,latexsym,rotate}
\usepackage[ps,arc,frame]{xy}
\usepackage{amsfonts,amssymb} 
\usepackage{psfrag}
\epsfverbosetrue
\usepackage{graphicx}

\newcommand{\rxyn}[2]{{\begin{xy} 0;<2mm,0mm>:<0mm,2mm>::0;0,
,(5,-2)*{a}
,(10,-1.8)*{b}
,(15,-2)*{c}
,(20,-2)*{d}
,(25,-2)*{e}
,(30,-2)*{f}
,(35,-2)*{g}
,(40,-2)*{h}
,(2,-5)*{a}
,(2,-10)*{b}
,(2,-15)*{c}
,(2,-20)*{d}
,(2,-25)*{e}
,(2,-30)*{f}
,(2,-35)*{g}
,(2,-40)*{h}
,(5,-5)*\cir(#1,0){}
,(10,-5)*\cir(#1,0){}
,(15,-5)*\cir(#1,0){}
,(20,-5)*\cir(#1,0){}
,(25,-5)*\cir(#1,0){}
,(30,-5)*\cir(#1,0){}
,(35,-5)*\cir(#1,0){}
,(40,-5)*\cir(#1,0){}
,(5,-10)*\cir(#1,0){}
,(10,-10)*\cir(#1,0){}
,(15,-10)*\cir(#1,0){}
,(20,-10)*\cir(#1,0){}
,(25,-10)*\cir(#1,0){}
,(30,-10)*\cir(#1,0){}
,(35,-10)*\cir(#1,0){}
,(40,-10)*\cir(#1,0){}
,(5,-15)*\cir(#1,0){}
,(10,-15)*\cir(#1,0){}
,(15,-15)*\cir(#1,0){}
,(20,-15)*\cir(#1,0){}
,(25,-15)*\cir(#1,0){}
,(30,-15)*\cir(#1,0){}
,(35,-15)*\cir(#1,0){}
,(40,-15)*\cir(#1,0){}
,(5,-20)*\cir(#1,0){}
,(10,-20)*\cir(#1,0){}
,(15,-20)*\cir(#1,0){}
,(20,-20)*\cir(#1,0){}
,(25,-20)*\cir(#1,0){}
,(30,-20)*\cir(#1,0){}
,(35,-20)*\cir(#1,0){}
,(40,-20)*\cir(#1,0){}
,(5,-25)*\cir(#1,0){}
,(10,-25)*\cir(#1,0){}
,(15,-25)*\cir(#1,0){}
,(20,-25)*\cir(#1,0){}
,(25,-25)*\cir(#1,0){}
,(30,-25)*\cir(#1,0){}
,(35,-25)*\cir(#1,0){}
,(40,-25)*\cir(#1,0){}
,(5,-30)*\cir(#1,0){}
,(10,-30)*\cir(#1,0){}
,(15,-30)*\cir(#1,0){}
,(20,-30)*\cir(#1,0){}
,(25,-30)*\cir(#1,0){}
,(30,-30)*\cir(#1,0){}
,(35,-30)*\cir(#1,0){}
,(40,-30)*\cir(#1,0){}
,(5,-30)*\cir(#1,0){}
,(10,-35)*\cir(#1,0){}
,(15,-35)*\cir(#1,0){}
,(20,-35)*\cir(#1,0){}
,(25,-35)*\cir(#1,0){}
,(30,-35)*\cir(#1,0){}
,(35,-35)*\cir(#1,0){}
,(40,-35)*\cir(#1,0){}
,(5,-30)*\cir(#1,0){}
,(10,-40)*\cir(#1,0){}
,(15,-40)*\cir(#1,0){}
,(20,-40)*\cir(#1,0){}
,(25,-40)*\cir(#1,0){}
,(30,-40)*\cir(#1,0){}
,(35,-40)*\cir(#1,0){}
,(40,-40)*\cir(#1,0){}
#2\end{xy}}}

\newcommand{\double}[1]{\mathbb{#1}}

\newcommand{\cc}{\double{C}}

\newcommand{\rr}{\double{R}}
\newcommand{\zz}{\double{Z}}

\newcommand{\aaa}{\mathcal{A}}

\newcommand{\uu}{\mathcal{U}}

\newcommand{\mm}{\mathcal{M}}

\newcommand{\pp}{\pmatrix}

\newcommand{\dd}{\mathcal{D}}

\newcommand{\llll}{\mathcal{L}}

\newcommand{\mmf}{\hbox{$^f$\hspace{-0.2cm} $\mathcal{M}$}}
\newcommand{\Tf}{\hbox{$^f$\hspace{-0.2cm} $T$}}

\newcommand{\ddf}{\hbox{$^f$\hspace{-0.15cm} $\mathcal{D}$}}

\newcommand{\ttt}{{\rm tr}}

\def\ddd{{\,\hbox{$\partial\!\!\!/$}}}

\newcommand{\dee}{\hbox{\rm{D}}}

\newcommand{\ot}{\otimes}
\newcommand{\op}{\oplus}

\newcommand{\bb}{\begin{eqnarray}}
\newcommand{\ee}{\end{eqnarray}}
\newcommand{\eee}{\nonumber\end{eqnarray}}

\begin{document}

\font\twelve=cmbx10 at 13pt
\font\eightrm=cmr8

\thispagestyle{empty}

\begin{center}

Institut f\"ur Mathematik  $^1$ \\ Universit\"at Potsdam
\\ Am Neuen Palais 10 \\14469 Potsdam \\ Germany\\

\vspace{2cm}

{\Large\textbf{The Inverse Seesaw Mechanism \\ in Noncommutative
Geometry}} \\

\vspace{1.5cm}

{\large Christoph A. Stephan$^{1,2}$}

\vspace{2cm}

{\large\textbf{Abstract}}
\end{center}
In this publication we will implement the inverse Seesaw mechanism
\cite{valle} into the noncommutative
framework  \cite{con} on the basis of the $AC$-extension of the Standard Model
\cite{beyond2}. 
The main difference to the classical $AC$ model
is the chiral nature of the $AC$ fermions with respect to
a $U(1)_X$ extension of the Standard Model gauge group.
It is this extension which allows us to couple the
right-handed neutrinos via a gauge invariant mass term
to left-handed $A$-particles. The natural scale
of these gauge invariant masses is of the order of  $10^{17}$ GeV
while the Dirac masses of the neutrino and the 
$AC$-particles are generated dynamically and are therefore
much smaller ($\sim 1$ GeV to $\sim  10^6$ GeV). From
this configuration a working inverse Seesaw mechanism
for the neutrinos is obtained.

\vspace{2cm}

\noindent
PACS-92: 11.15 Gauge field theories\\
MSC-91: 81T13 Yang-Mills and other gauge theories

\vskip 1truecm

\noindent \\

\vspace{1.5cm}
\noindent $^2$ christophstephan@gmx.de\\

\section{Introduction}

We present an extension of the Standard Model in
its noncommutative formulation \cite{con} which 
implements the inverse Seesaw mechanism
\cite{valle}. As previous extensions of the
Standard Model within the noncommutative
framework,  \cite{beyond2,vector,newcolour}
and \cite{newscalar}, this
model is based on the classification of finite
spectral triples \cite{1,2,3,Spinlift,4,5}. It is a variant of
the $AC$-model found in \cite{beyond2} which contains
the Standard Model fermions as well
as a two new species of particles, the $A$- and 
the $C$-particles. In this paper  we work with
a spectral triple
where the $KO$-dimension  of the internal part
 taken to be six \cite{barrett,cc}.
In the original $AC$-model, \cite{beyond2}, the new particles 
were electrically charged with twice the electron charge
and turned out to be viable candidates for dark
matter \cite{khlop}. Here we will assume them
to posses chiral charges of a new $U(1)_X$ gauge group.
\\
An open problem in noncommutative geometry is 
the realisation of the mass mechanism for neutrinos.
The spectral action implies
not only the Standard Model action and the Einstein-Hilbert action
but also a set of conditions imposed on the couplings of the bosonic
and fermionic fields.
One finds \cite{thum,cc} that the condition imposed on the Yukawa couplings 
demands additionally to the top quark coupling at least a second
Yukawa coupling to be of order one. This fact strongly suggests
a Seesaw-like mechanism where the additional large Yukawa
coupling is taken to be one of the neutrino couplings.
\\
While in KO-dimension zero the neutrino masses are of Dirac 
type \cite{gracia,neutrino,ra2}, KO-dimension six also allows for 
Majorana masses \cite{barrett,cc} and the Seesaw mechanism.
The price to be paid for the Majorana mass is a violation of
one of the axioms of noncommutative geometry, namely the axiom of 
orientability \cite{ko6}. This problem can be overcome by 
introducing a second layer for the internal algebra  in the finite
part of the  spectral triple \cite{cc,con}, or by a modification of the 
spectral action \cite{sit}.
\\
Nevertheless a numerical analysis of the Standard Model with SeeSaw mechanism
\cite{cc,sm1,sm2}  shows that at least one the gauge invariant Seesaw masses is
of the order of
$\sim 10^{14}$ GeV, while one would expect the elements of 
the Majorana mass matrix to be of the
order of $\sim 10^{17}$ GeV, the cut-off scale of the spectral action. 
The inverse Seesaw shifts this discrepancy to an intermediate energy scale
associated to the  vacuum expectation value of a new
scalar field. Furthermore the present
version of the $AC$-model is completely compatible with
the axioms of noncommutative geometry \cite{con} and 
can in principle be employed in models based on spectral
triples with  KO-dimension six and zero
(although a specific model for KO-dimension zero does
not exist, yet).
\\
The model presented here has as gauge group $G=U(1)_Y \times
SU(2) \times SU(3) \times U(1)_X$ where the Standard 
Model subgroup $G_{SM}=U(1)_Y \times
SU(2) \times SU(3)$
is broken by the usual Higgs mechanism to $U(1)_{em} \times SU(3)$
and the new subgroup $U(1)_X$  is broken to $\zz_2$
by a decoupled Higgs mechanism associated to a new 
scalar field. It is therefore the second extension of
the Standard Model within the noncommutative 
framework after \cite{newscalar} which has an enlarged scalar 
sector.
This new scalar field  generates the masses in the $AC$-sector.
Previous attempts to extend the Standard Model within the
framework of noncommutative geometry proved 
to be extremely difficult. Most of the early attempts 
unfortunately failed to produce physically interesting
models \cite{beyond}. 
\\
It would of course also be desirable to gain a deeper
understanding of the origin
of the internal space, i.e. the source of the matrix algebra.
There are hints that a connection to loop quantum gravity
exists \cite{dan}. Also  double Fell bundles
seem a plausible structure in noncommutative geometry \cite{ra1}. 
They could provide a deep connection to category theory
and give better insights into the mathematical structure of
almost-commutative geometries such as the Standard Model.
\\
This paper is organised as follows: In section two we give the construction
of the internal space based on a minimal Krajewski diagram. 
We calculate the lift of the gauge group and
the fluctuated Dirac operator. This fluctuation leads to the 
Standard Model Higgs and a new scalar field.
We then calculate the relevant parts of the
spectral action which provides the potential
for the Higgs and the  scalar field, the new parts in
the fermionic Lagrangian, as well as constraints
on the quadratic couplings and the quartic couplings
of the scalar fields, the  Yukawa couplings of the fermions 
and  the gauge couplings of the non-abelian subgroup
of the gauge group. 
\\
In section three we analyse the mass matrix for
the neutrino sector coupled to the $AC$-sector 
and calculate the mass eigenvalues for the
light mass eigenstates and the  heavy mass eigenstates.
We then estimate the scale of the vacuum expectation
value of the  new scalar field and give also a
few estimations of the range of masses
in the $AC$-sector.

\section{The model}

The model proposed in this paper is a variant of
the $AC$-model \cite{beyond2}, but with a different charge
assignment for the $U(1)$ subgroups of the
gauge group. The internal Hilbert space which
encodes the multiplet structure of the gauge group 
consists of the usual chiral Standard Model multiplets (i.e.
six quarks as $SU(2)$ doublets/singlets and $SU(3)$
triplets and six leptons as $SU(2)$ doublets/singlets)
plus $A$ and $C$ particles being $SU(2)$ and $SU(3)$
singlets. 
\\
Internal spaces of almost-commutative geometries are
conveniently encoded in Krajewski diagrams \cite{kraj}.
The Krajewski diagram for this model is depicted in figure 1. 
Note that the allowed  mass term connecting right-handed $C$-particles
and to right-handed $A$-antiparticles does not appear explicitly
since we have left out the antiparticles to keep the
diagram simple. This Krajewski diagram is based on a minimal 
diagram that can be obtained by deleting the arrow for
the right-handed neutrino.
\begin{figure}
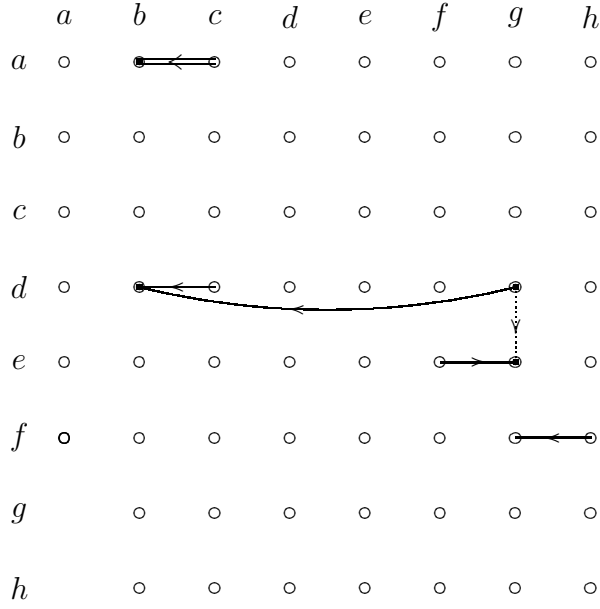

\begin{center}
\begin{tabular}{c}
\rxyn{0.4}{
,(10,-5);(15,-5)**\dir2{-}?(.4)*\dir2{<}
,(10,-5)*\cir(0.2,0){}*\frm{*}
,(10,-20);(15,-20)**\dir{-}?(.4)*\dir{<}
,(10,-20);(35,-20)**\crv{(22.5,-23)}?(.4)*\dir{<}
,(10,-20)*\cir(0.2,0){}*\frm{*}
,(30,-25);(35,-25)**\dir{-}?(.6)*\dir{>}
,(35,-30);(40,-30)**\dir{-}?(.4)*\dir{<}
,(35,-20);(35,-25)**\dir{..}?(.6)*\dir{>}
,(35,-20)*\cir(0.2,0){}*\frm{*}
,(35,-25)*\cir(0.2,0){}*\frm{*}
}
\end{tabular}
\caption{Krajewski diagram of the extended Standard Model. The
dotted  line indicates  the gauge invariant mass term connecting
the right-handed neutrino to the left-handed  $A$-particle.}
\end{center}
\end{figure}
\\
As matrix  algebra for the internal space we choose 
$\aaa = \cc \op M_2(\cc) \op M_3(\cc) \op \cc \op \cc \op \cc
\op \cc \op \cc$. This is exactly the $AC$-model
algebra \cite{beyond2} in KO-dimension six. 
From the Krajewski diagram  we read off the representation
for $\aaa\ni (a,b,c,d,e,f,g,h)$:
\bb 
\rho _L =\pp{b\ot 1_3&0&0 &0 \cr  0& b & 0 &0 \cr 0 & 0& g &0
\cr 0& 0& 0& \bar{g} },&&
\rho _R=\pp{c\ot 1_3&0&0&0&0&0  \cr  0&\bar{c} \ot 1_3&0&0&0 &0 
\cr 0&0& \bar{c} &0&0 &0  \cr 0&0&0&g &0 &0  \cr 0&0&0&0& e&0
 \cr 0&0&0&0& 0& h  },\cr \cr \cr
\rho _L^c=\pp{1_2 \ot  a&0 &0 & 0 \cr  0&d\, 1_2  & 0 & 0 \cr 0&0&  e & 0 
\cr  0& 0& 0 & f },&&
\rho _R^c=\pp{ a &0&0&0 &0 & 0 \cr  0& a &0&0 &0 & 0
\cr 0&0&  d &0 &0 & 0 \cr 0&0&0& d &0  & 0 \cr 0&0&0& 0&  e & 0
\cr & 0& 0& 0& 0 &f},
\label{representation}
\ee
The internal part $\dd$ of the Dirac operator can be decomposed
as follows:
\bb
\dd = \pp{\Delta & T \cr \bar{T} & \bar{\Delta}}, \quad  {\rm with} \quad
\Delta = \pp{ 0 & \mm \cr \mm^* & 0}, 
\label{diracop}
\ee
where the submatrix $\mm$ is given by 
\bb 
\mm=\pp{(M_u, M_d)\ot 1_3 & 0 &0&0\cr  0&(M_e, M_\nu) &0 &0
\cr 0&  (0,M_{\nu A}) & M_A &0 \cr 0& 0& 0&M_C },\cr \cr \cr 
\label{massmatrix}
\ee
Here 
$ (M_u, M_d)$ is the mass matrix of the quarks, $(M_e, M_\nu)$ is the
the mass matrix of the leptons and $M_A$ $(M_C)$ is the mass matrix of
the $A$-particles  $(C-$particles). 
The gauge invariant mass matrix connecting the right-handed
neutrinos to the left-handed $A$-particles is $(0,M_{\nu A})$.
The submatrix $T$ is
\bb
T = \pp{0  &0 &0 &0 \cr
0  &0  &  0& M_{AC} \cr
0&0  & M_{AC}^t & 0},
\ee
with $M_{AC}$ a Majorana-type mass matrix connecting 
right-handed $C$-particles to right-handed $A$-antiparticles.
It is not a gauge invariant mass term and will be associated to
the new scalar field.
We assume that the $A$- and $C$-particles
come, as the Standard Model particles in three generations, so  $M_{\nu A},
M_A,M_C,M_{AC}  \in M_3(\cc)$.
\\
The non-abelian subgroup of unitaries of the matrix algebra $\aaa$ is 
$\uu^{nc} = U(2) \times U(3)$. It contains two $U(1)$ subgroups
via the determinant that
may be lifted by central extensions to the fermionic Hilbert space \cite{fare}. 
We will call these two subgroups $\det(U(2) )= U(1)_Y$ and $\det(U(3))= U(1)_X$. 
The first one is nothing else but the Standard Model hypercharge
subgroup and the second one is associated with the $AC$-particles.
The $AC$-particles are neutral with respect to the 
Standard Model gauge group, i.e. the $AC$-particles
are $SU(2) \times SU(3)$ singlets and have zero hypercharge.
On the other hand the Standard Model particles are 
neutral with respect to the $U(1)_X$.
It follows that the gauge group of our model is 
$G = U(1)_Y \times SU(2) \times SU(3) \times U(1)_X$.
\\
An anomaly free lift of $\uu^{nc}$ to the Hilbert space
is achieved by the following central charge assignment,
normalised to unity, of
$A$ and $C$ for the $U(1)_X$ subgroup:
\begin{center}
\begin{tabular}{c|c|c|c|c}
& $A_L$ & $A_R$ & $C_L$ & $C_R$ \\
\hline
$Q_X$  & 0 & 1 & 1 & 0
\end{tabular}
\end{center}
It is remarkable that the representation (\ref{representation}) 
allows for a charge assignment which produces a new 
particle sector sterile to the Standard Model gauge group,
yet chiral under the new $U(1)_X$ subgroup while at
the same time  allowing for the gauge invariant mass term 
$M_{\nu A}$.
\\
The anomaly free lift $L$ decomposes into the usual 
Standard Model lift $L_{SM}$ which can be found in \cite{fare}
and the lift  $L_X$ acting on the $AC$-particles. This can be written
as
\bb
L(\det(u),\det(v),u,v) = L_{SM}( \det(v), u,v) \op L_X(\det(u))
\ee
where $u\in U(2)$, $v \in U(3)$.
For the new part of the lift $L_X$ we find
\bb
L_X(\det(u)) = {\rm diag}(1,\det(u)^{1},\det(u)^{1},1;1,\det(u)^{-1},\det(u)^{-1},1).
\ee
The semicolon
divides the particles from the antiparticles and the $U(1)_X$-charges of $A$ and
$C$  have been used. 
\\
Next we need to fluctuate the Dirac operator \cite{con}  to obtain the gauge 
bosons as well as the Higgs field $\phi$  and the new scalar field 
$\varphi$.
We define the fluctuated Dirac operator $\ddf$ according to \cite{1}:
\bb
\ddf = \sum_i r_i L(\det(u_i),\det(v_i),u_i,v_i) \dd L(\det(u_i),\det(v_i),u_i,v_i)^{-1},
\;\; r_i \in \rr .
\ee
One obtains the standard Higgs doublet $\phi$ embedded into a quaternion
and a new complex scalar field $\varphi$ due to the fact that the lift does not
commute with the Dirac operator $\dd$. The only part of the mass matrix
$ \mm$ of $\dd$ commuting with the lift is $M_{\nu A}$ which is therefore
a gauge invariant mass.
We find for the fluctuated mass matrices 
\bb
\mmf = \pp{\phi (M_u, M_d)\ot 1_3 & 0 &0&0\cr  0&\phi (M_e, M_\nu) &0 &0
\cr 0&  (0,M_{\nu A}) & \varphi M_A &0 \cr 0& 0& 0& \varphi M_C }
\ee
and
\bb
\Tf = \pp{0&0  &0 &0 \cr
0   &  0 &0 & \bar{\varphi} M_{AC} \cr
0&0 & \bar{\varphi} M_{AC}^t & 0},
\ee
with $ \varphi = \sum_i r_i \det(u_i)^{-1}$. The new scalar field
$\varphi$ is also neutral with respect to the Standard Model gauge group
and has $U(1)_X$-charge $Q_X = -1$.
From these mass matrices we can calculate the spectral action
which will give us the kinetic term of the scalars as well
as the potential for the Higgs field and the new scalar field. 
\\
According to \cite{con} the spectral action $S_{CC}$ is given by 
the number of eigenvalues of the Dirac operator $\dee$
up to a cut-off energy $\Lambda$.  $\dee = \ddd \otimes
1_{dim. \mathcal{H}_f} + \gamma^5 \otimes \dd$ is
the Dirac operator of the full almost-commutative geometry.
The spectral action can be
written approximately with help of a positive
cut-off function $f$ and then be calculated asymptotically 
via a heat-kernel expansion:
\bb
S_{CC} = \ttt ( f \left(\frac{\dee^2}{\Lambda^2}\right) ) = \frac{1}{16 \pi^2}
\, \int {\rm dV} ( a_4 f_4 \Lambda^4 + a_2 f_2 \Lambda^2 + a_0 f_0 + o(\Lambda^{-2})
)
\ee
Here $f_i$ are the first moments of the cut-off function $f$. They enter as
free parameters into the model. The heat-kernel coefficients $a_i$ are well
known \cite{gilk} and for the present calculation only $a_2$ and 
$a_0$ will be of concern. Note that we use the numerating convention
of \cite{cc}, where the number of the coefficient $a_i$ corresponds to
the power of $\Lambda$.
\\
The coefficient $a_2$ will give us the mass terms of the potential for
the scalar fields while $a_0$ will provide for the kinetic terms for the 
scalar fields 
the quartic couplings of the potential and also mass terms.
All the following relations
hold at the cut-off energy $\Lambda$ and are not stable
under the renormalisation group flow. 
\\
To calculate the relevant parts of  
$a_2$ and $a_0$ we need the traces of $\ddf^2$ and
$\ddf^4$. These calculations are very similar to 
those presented in detail in \cite{newscalar}. Therefore
we will only present the final results.
\\
One observes that the scalar fields so far have mass dimension zero.
We have to normalise the scalar fields   $\phi (M_u,M_d) 
\rightarrow  \tilde \phi (\mathcal{Y},\mathcal{Y})$
and $\varphi M \rightarrow \tilde \varphi \mathcal{Y}$  to obtain the standard kinetic
terms of the Lagrangian. Here $(\mathcal{Y},\mathcal{Y})$ and $ \mathcal{Y}$
are the Yukawa coupling matrices of the quarks/leptons and the $AC$-particles. It 
is very convenient to immediately drop the $\tilde \phi / \tilde \varphi$ notation 
for the normalised scalar fields and return to $\phi / \varphi$ since only
the normalised fields appear from now on.
\\
For the real scalar fields $\phi_i$ and $\varphi_j$ 
we use  the standard normalisation
\bb
 \phi = \frac{1}{\sqrt{2}} \pp{ \phi_1 +i \phi_2 \cr \phi_3+i \phi_4}
\quad {\rm and} \quad 
 \varphi = \frac{1}{\sqrt{2}} ( \varphi_1 +i \varphi_2 ).
\ee
All the Standard Model fields acquire their well known Standard Model
Lagrangian, for details see \cite{cc}. For example the 
Higgs field Lagrangian is 
\bb
\llll_{Higgs}&&=  (D_\mu  \phi)^* (D^\mu  \phi)
+ \mu_1^2 | \phi|^2 - \frac{\lambda_1}{6} \, 
| \phi|^4.
\label{kinetic}
\ee
For the new scalar $\varphi$ we find the following standard Lagrangian
\bb
\llll_{scalar}&&=  (D_\mu  \varphi)^* (D^\mu  \varphi)
+ \mu_2^2 | \varphi|^2 - \frac{\lambda_2}{6} \, 
| \varphi|^4.
\label{kinetic}
\ee
The potential for $\varphi$ will have a nontrivial minimum, just as
the Higgs potential. It follows that the $U(1)_X$ subgroup 
is broken dynamically. 
Notice that there is no term proportional to $ | \phi|^2 | \varphi|^2$
mixing the Standard Model Higgs $\phi$ with the new scalar $\varphi$
as it was the case in the model presented in \cite{newscalar}. We also
get the standard Lagrangian $\llll_{X} = - 1/4 \, F_{X,\mu \nu} \, F^{\mu \nu}_X$
for the new subgroup
$U(1)_X$ with coupling $g_4$ as well a conformal coupling of
the new scalar $\varphi$ to the curvature scalar $R$. These terms 
will not concern us in the following.
\\
The symmetry breaking pattern of the gauge group $G$ is 
\bb
U(1)_Y \times SU(2) \times SU(3) \times U(1)_X
\rightarrow U(1)_{ew} \times SU(3).
\ee
The fermionic action $(\psi, \ddf \psi)$, where $(\cdot,\cdot)$ denotes the
scalar product on the Hilbert space, provides for the mass terms 
of the model. Apart from the Yukawa terms of the Standard Model
we have the following terms in the Lagrangian:
\bb
\llll_{Yukawa}&=& (\nu_R, M_{\nu A} \, A_L) + (A_L, \varphi  \, \mathcal{Y}_A \, A_R)
+ (C_L, \varphi  \, \mathcal{Y}_C \, A_C)
\nonumber \\
&& +  (A_R,\bar \varphi  \, \mathcal{Y}_{AC} \, C_R)+ {\rm c.c.}
\label{yukawa}
\ee
It will be the gauge invariant  mass term $ (\nu_R, M_{\nu A} \, A_L)$, connecting the
right-handed neutrinos to the left-handed $A$-particles, which is
responsible for the inverse Seesaw mechanism.
The vacuum expectation value (vev) of the scalar field $\varphi$ 
is a free parameter of the model. Later we will estimate the vev of
$\varphi$ by using
the requirement that the entries of $M_{\nu A}$ should be of
the order of the cut-off $\Lambda$ in the spectral action.
\\
For all the parameters in the Lagrangian the spectral action provides
a set of constraints \cite{thum,cc}.
Let us first regard the  constraints on the dimensionful and dimensionless 
couplings of the Higgs and the new scalar field. For the quadratic couplings
we find:
\bb
\mu_1^2 = 2 \frac{f_2}{f_0} \Lambda^2 - 2 \frac{\ttt(\mathcal{Y}_\nu^* \mathcal{Y}_\nu
M_{\nu A}^* M_{\nu A}) }{Y_2} 
\quad {\rm and} \quad 
\mu_2^2= 2 \frac{f_2}{f_0} \Lambda^2 -\frac{\ttt(\mathcal{Y}_A^* \mathcal{Y}_A
M_{\nu A}^* M_{\nu A}) }{\tilde Y_2} ,
\ee
where as usual $Y_2= \ttt(\mathcal{Y}_u^* \mathcal{Y}_u 
+\mathcal{Y}_d^* \mathcal{Y}_d +\mathcal{Y}_e^* \mathcal{Y}_e
+\mathcal{Y}_\nu^* \mathcal{Y}_\nu  )$ is the trace of the Standard Model
Yukawa matrices squared and $\tilde Y_2 = \ttt (\mathcal{Y}_A^* \mathcal{Y}_A+\mathcal{Y}_C^* \mathcal{Y}_C+\mathcal{Y}_{AC}^* \mathcal{Y}_{AC} )$ is
its analogue for the $AC$-sector. 
\\
Since $M_{\nu A}$ is gauge invariant the entries of the
matrix should be of the order of the cut-off scale, so $(M_{\nu A})_{ij}
\sim \Lambda$. 
We observe that   $\ttt (\mathcal{Y}_\nu^* \mathcal{Y}_\nu
M_{\nu A}^* M_{\nu A}) \sim \ttt(\mathcal{Y}_A^* \mathcal{Y}_A
M_{\nu A}^* M_{\nu A}) \sim \Lambda^2$ allows us to decouple the 
vacuum expectation values of $\phi$ and $\varphi$ from the cut-off
scale $\Lambda$. This means especially that the $W$-mass is
decoupled from the cut-off scale.
\\
For  the dimensionless quartic couplings we find:
\bb
\lambda_1 = 24 \frac{\pi^2}{f_0} \frac{H}{Y_2^2},
\quad \lambda_2 =24 \frac{\pi^2}{f_0} \frac{\tilde H}{\tilde Y_2^2}, 
\ee
with $H= \ttt((\mathcal{Y}_u^* \mathcal{Y}_u)^2 
+(\mathcal{Y}_d^* \mathcal{Y}_d)^2 +(\mathcal{Y}_e^* \mathcal{Y}_e)^2
+(\mathcal{Y}_\nu^* \mathcal{Y}_\nu)^2  )$ 
and $\tilde H = \ttt ((\mathcal{Y}_A^* \mathcal{Y}_A)^2+(\mathcal{Y}_C^* \mathcal{Y}_C)^2+(\mathcal{Y}_{AC}^* \mathcal{Y}_{AC})^2 )$.
\\
The constraints for the Yukawa couplings are $Y_2 = \tilde Y_2 = 4\, \pi^2 / f_0$
which, together with $g_3^2 = g_2^2 = \pi^2 / f_0$ gives the
final set of constraints for the dimensionless couplings at the 
cut-off $\Lambda$:
\bb
g_2^2=g_3^2 = \frac{\lambda_1}{24} \, \frac{Y_2^2}{H}
= \frac{\lambda_2}{24} \, \frac{\tilde Y_2^2}{\tilde H} =
\frac{Y_2}{4} = \frac{\tilde Y_2}{4},
\label{constraints}
\ee
where $g_2$ is the $SU(2)$ coupling and $g_3$ is the
$SU(3)$ coupling.
\\
From these relations it is now possible to deduce the 
cut-off scale $\Lambda$ using renormalisation
group techniques \cite{cc}. From the constraint
$g_2 = g_3$ follows $\Lambda = 1.1 \times 10^{17}$ GeV.
With $\lambda_1 =  24 g_2^2 \, H / Y_2^2$ at $\Lambda$ follows
then a Higgs mass of the  order of $170$ GeV \cite{cc}.
The value of the Higgs mass may change considerably
for different models beyond the Standard Model \cite{vector,
newcolour,newscalar}. With the same procedure 
we could also calculate the low energy value of $\lambda_2$.
But since only the overall scale of the $\varphi$-vev
matters here, we will postpone these investigations
to a later publication.

\section{The inverse Seesaw Mechanism}

The relevant mass terms for the inverse Seesaw mechanism
are the Dirac masses $m_\nu$  of the neutrinos, the Dirac masses
$m_A$ of the $A$-particles and the gauge invariant mass term
$M_{\nu A}$ connecting the right-handed neutrinos to the
left-handed $A$-particles. 
To simplify the model a bit, we will assume that the Dirac masses
of the $C$-particles, as well as the mass terms connecting $A$-particles and
$C$-antiparticles, are much smaller than the $A$-particle masses.
\\
It follows from the constraints (\ref{constraints}) that the neutrino 
Yukawa couplings should be of order one, so the Dirac masses
are $m_\nu \sim 100$ GeV. We assume that the same holds for the Yukawa couplings
of the $A$-particles, therefore their masses should be of the order
of the vacuum expectation value of the scalar field $\varphi$, which
is a free parameter.
\\
From the Lagrangian $\llll_{Yukawa}$ (\ref{yukawa}) we can
deduce the following mass matrix for the neutrinos and the
$A$-particles:
\bb
M = \pp{0 & m_\nu & 0 & 0 \cr m_\nu & 0 & M_{\nu A} & 0 \cr 
0& M_{\nu A} &0 & m_A  \cr 0&0&m_A &0 }.
\ee
This type of matrix is well known \cite{valle}. 
To transparently calculate the eigenvalues of $M$ we will 
take $m_\nu, M_{\nu A}$ and $m_A$ to be
the Dirac masses of the first Standard Model + $AC$
family. 
\\
As was shown in \cite{valle} the eigenvalues
of  $M$ are given by
\bb
m_{1/2}^2 =  m_\nu^2 \frac{m_A^2}{M_{\nu A}^2} \quad {\rm and}
\quad m_{3/4}^2 =  M_{\nu A}^2.
\ee
One sees immediately that one obtains two the light mass eigenstates
and two heavy mass eigenstates $M_{\nu A} \sim \Lambda = 1.1 \times
10^{17}$ GeV. 
\\
Assuming that the light  mass eigenstates corresponds to
the light neutrino, which we assume to be  of the order of
1 eV we can deduce that the approximate mass scale for the 
vacuum expectation value of the new scalar field $\varphi$
should be of the order of $10^6$ GeV.
\\
Speculating further we may assume that the variation of the masses
in the $AC$ sector is similar to the 
Standard model, i.e. 
\bb
\frac{m_C}{ m_A}  \sim \frac{m_e}{m_{top}} \sim 10^{-6},
\ee
where $m_C$ is a generic $C$-particle mass, $m_e$ is
the electron mass and $m_{top}$ the top quark mass. 
This would imply that the mass of the lightest $C$-particle could
be as low as $m_C \sim m_A \times 10^{-6} \sim 1$ GeV.
The mass of the $X$ boson would be $m_{X} \sim \sqrt{g_4} \,
10^6$ GeV, where the $U(1)_X$ gauge coupling $g_4$ is 
essentially a free parameter.

\section{Conclusion and outlook}

We have shown in this publication that the noncommutative
framework  \cite{con} allows us
to successfully implement the inverse Seesaw mechanism
\cite{valle} on the basis of the $AC$-extension of the Standard Model
\cite{beyond2}. The main difference to the classical $AC$ model
is the chiral nature of the $AC$ fermions with respect to
a $U(1)_X$ extension of the Standard Model gauge group.
It is this extension which allows us to couple the
right-handed neutrinos via a gauge invariant mass term
to left-handed $A$-particles. Since the natural scale
of these gauge invariant masses is of the order of 
the cut-off scale of the spectral action ($\sim 10^{17}$ GeV),
while the Dirac masses of the neutrino and the 
$AC$-particles are generated dynamically and therefore
much smaller ($\sim 1$ GeV to $\sim  10^6$ GeV), a
working inverse Seesaw mechanism is obtained.
In the usual realisation of the Seesaw mechanism 
at least one of the Majorana mass terms has to be several
orders of magnitude smaller than the cut-off scale ($M_M
\sim 10^{14}$ GeV, \cite{sm1,sm2}).
\\
Compared to the usual Seesaw mechanism \cite{barrett,con}
there are several differences, some of which could turn out to
be advantageous:
\begin{itemize}
\item the spectral triple on which the model is based 
fulfils all axioms of noncommutative geometry \cite{con}.
\item there is no principle obstacle to realise this or
a similar mechanism in KO-dimension zero.
\item the model does not allow neutrinoless double-$\beta$-decay.
So it is  falsifiable by experiment.
\item the $AC$-particles could produce stable or sufficiently
long lived particles that could serve as dark matter. 
\item since the $A$-particle mixes with the Standard Model
neutrino its Dirac mass matrix as well as the mass matrix
connecting the two particles introduce new CP violating
phases. This may be interesting for leptogenesis.
\end{itemize}
An inconvenience of the model is the Higgs mass of the
order of $170$ GeV. This mass range is now practically
excluded by the Tevatron \cite{tevatron}. It will be
interesting to see if the inverse Seesaw mechanism
can be implemented into other models beyond the
Standard Model within the framework of noncommutative
geometry. These models usually provide stronger
constraints on the coupling constants, 
\cite{beyond2,vector,newcolour} and \cite{newscalar},
and may therefore be very predictive.

\subsection*{Acknowledgements}

The author gratefully acknowledges the funding of his work
by the Deutsche Forschungsgemeinschaft.

\vfil\eject
\enlargethispage{1cm}
\thispagestyle{empty}

\end{document}